\documentclass{PoS}
\usepackage{amsmath}
\usepackage{mwe}

\title{Lattice Calculation of the Proton Charge Radius}

\ShortTitle{Proton Charge Radius}

\author{{William Detmold} \\
Massachusetts Institute of Technology \\
E-mail: \email{wdetmold@mit.edu}}

\author{\speaker{Anthony Grebe}\\
        Massachusetts Institute of Technology \\
        E-mail: \email{agrebe@mit.edu}}
        
\author{{Phiala Shanahan} \\
Massachusetts Institute of Technology \\
E-mail: \email{pshana@mit.edu}}

\abstract{
The charge radius of the proton has been measured in scattering and spectroscopy experiments using both electronic and muonic probes. The electronic and muonic measurements are discrepant at $5\sigma$, giving rise to what is known as the proton radius puzzle.

With the goal of resolving this, we introduce a novel method of using lattice QCD to determine the isovector charge radius -- defined as the slope of the electric form factor at zero four-momentum transfer -- by introducing a mass splitting between the up and down quarks.  This allows us to access timelike four-momentum transfers as well as spacelike ones, leading to potentially higher accuracy in determining the form factor slope at $Q^2 = 0$ by interpolation.  In this preliminary study, we find a Dirac isovector radius squared of $0.320 \pm 0.074$ fm$^2$ at quark masses corresponding to $m_\pi = 450$ MeV.  We compare the feasibility of this method with other approaches of determining the proton charge radius from lattice QCD.

}

\FullConference{The 36th Annual International Symposium on Lattice Field Theory - LATTICE2018\\
		22-28 July, 2018\\
		Michigan State University, East Lansing, Michigan, USA.}

\begin{document}

\section{Introduction}
The proton charge radius has been determined experimentally through both scattering and spectroscopy experiments.  Scattering experiments measure the differential cross section for lepton-proton scattering at discrete values of angle and energy; spectroscopy experiments measure corrections to the Lamb shift due to the finite size of the proton.  These experiments can be performed with either electrons or muons as the probes, and the Standard Model predicts that the measurement of the proton charge radius should be independent of the probe.  However, electronic measurements (scattering and spectroscopy combined) give a proton charge radius of 0.8775 $\pm$ 0.0051 fm \cite{crema}, whereas muonic spectroscopy gives a radius of 0.84087 $\pm$ 0.00039 fm \cite{codata}.  This unresolved five-sigma discrepancy, called the proton radius puzzle, potentially indicates physics beyond the Standard Model.

Experimentally, elastic scattering measurements at 4-momentum transfer $q^2=-Q^2$ determine the electric and magnetic form factors $G_{E,M} (Q^2)$.  The proton charge radius is defined by the slope of the electric form factor at $Q^2 = 0$ \cite{thomson}:

\begin{equation}
\langle r_E^2 \rangle = -6 \left. \frac{dG_E (Q^2)}{d Q^2} \right|_{Q^2 = 0} .
\end{equation}

On the lattice, we can compute the hadronic matrix elements that enter into these scattering amplitudes.  The relevant matrix elements can be determined from three-point functions of the form
\begin{equation} \langle 0 | B_\alpha (t, \mathbf p_f) \mathcal O_\mu(\tau) \bar B_\beta(0, \mathbf p_i)| 0 \rangle , \end{equation}
where $B_\alpha(t,\mathbf p)$ is an interpolating operator for a state with momentum $\mathbf p$.  A lattice vector current $\mathcal{O}_\mu (\tau) = \sum_{\mathbf x} \bar \psi(\mathbf x,\tau) \gamma_\mu \psi (\mathbf x,\tau)$ is inserted at Euclidean time $\tau < t$.  When evaluated between proton spinors, $\mathcal O_\mu$ has the Euclidean-space decomposition \cite{peskin}
\begin{equation} \langle p_f | \mathcal O_\mu | p_i \rangle = \bar u (p_f) \left[ \gamma_\mu F_1(Q^2) + \frac{\sigma_{\mu\nu}q^\nu}{2m_p} F_2(Q^2) \right] u (p_i) . \end{equation} 
The Dirac and Pauli form factors $F_1, F_2$ are related to the electric and magnetic form factors by
\begin{align}
G_E (Q^2) &= F_1 (Q^2) - \frac{Q^2}{4m_p^2} F_2 (Q^2) , \\
G_M (Q^2) &= F_1 (Q^2) + F_2 (Q^2) .
\end{align}
By respectively adjusting $\mathbf p_f$ and $\mathbf p_i$, we can obtain $G_E$ at various values of $Q^2$, from which the slope at $Q^2=0$, which defines the charge radius, can be determined.

Note that the charge radius depends only on the slope of $F_1$ and the value of $F_2$ at $Q^2 = 0$:
\begin{equation} \langle r^2 \rangle = -6 \left. \frac{dG_E (Q^2)}{d Q^2} \right|_{Q^2 = 0}
 = -6 \left. \frac{dF_1 (Q^2)}{d Q^2} \right|_{Q^2 = 0} + \frac{6}{4m_p^2} F_2 (0) . \end{equation}
$F_2 (0) = \mu_p - 1 = 1.7928473508(85)$, the anomalous magnetic moment of the proton, is known experimentally to accuracy better than one part in $10^9$ \cite{pdg}.  Thus, following the approach in \cite{andrew-1}, we need only compute the Dirac charge radius, $\langle r_1^2 \rangle =  -6 \left. \frac{dF_1 (Q^2)}{d Q^2} \right|_{Q^2 = 0} $.

\subsection{Disconnected Diagrams}
The electromagnetic current insertion $\mathcal O_\mu = \frac{2}{3}\bar u\gamma_\mu u - \frac{1}{3} \bar d \gamma_\mu  d$ includes so-called disconnected diagrams where the photon interacts with a  sea quark (which connects to the valence quarks in the proton only through gluon exchange).  Such diagrams are computationally difficult to evaluate.  To avoid this difficulty, we instead use the isovector current $\bar u \gamma_\mu u - \bar d \gamma_\mu d$.  In the isospin limit (where $m_u = m_d$, well approximated by physical reality), the disconnected diagrams cancel.  In this limit, we can use isospin rotations to replace the above current with $\bar d \gamma_\mu u$, which turns an up quark into a down quark and thus a proton into a neutron.

In the isospin limit, $\langle p | \bar u \gamma_\mu u - \bar d \gamma_\mu d | p \rangle = \langle p | \mathcal O_\mu | p \rangle - \langle n | \mathcal O_\mu | n \rangle$ (with $\mathcal O_\mu = \frac{2}{3} \bar u \gamma_\mu u - \frac{1}{3} \bar d \gamma_\mu d$), so all isovector form factors and squared radii are just the differences of the corresponding proton and neutron quantities.  The neutron charge radius squared has been measured to be $-0.1161 \pm 0.0022$ fm$^2$ \cite{pdg,neutron}, so the proton charge radius can be computed from this and the isovector radius.  Most lattice computations (e.g.~\cite{andrew-1,andrew-2}) use this approach due to its large reduction in computational cost.

\subsection{Momentum Quantization}
In a lattice computation with periodic spatial boundaries, 3-momentum is quantized in multiples of $2\pi/L$, where $L$ is the spatial extent of the lattice.  We could calculate the slope of $F_1$ at $Q^2 = 0$ using a finite difference (i.e.~$[F_1(Q^2)-F_1(0)]/Q^2$), but the momentum quantization makes this difference relatively large for currently realistic lattice sizes, leading to large momentum quantization effects.  In principle, we could control this by increasing $L$, but for the requisite volumes this is computationally infeasible.  
The restriction $Q^2 \geq 0$ for spacelike momentum transfers restricts us to an asymmetric difference when estimating the slope, further increasing the difficulty.

Most lattice determinations of the proton charge radius (e.g.~\cite{andrew-1,andrew-2}) calculate the form factor over a range of $Q^2$ and then fit a curve to the data.  This introduces model dependence whose significance can be difficult to quantify.  Thus, it would be advantageous to perform a direct determination of the slope using small values of $Q^2$, ideally by using a symmetric difference about $Q^2 = 0$.

\section{Our Method}
We can determine the form factors at negative $Q^2$ values, including $Q^2$ values arbitrarily close to 0, if we break isospin symmetry by using non-degenerate quark masses.  Then the current $\bar d \gamma_\mu u$ changes nucleon energy even when $\mathbf q \equiv \Delta \mathbf p = 0$.  Such a current gives $Q^2 = \mathbf q^2 - (\Delta E)^2$, which can be negative and arbitrarily close to 0 as $\delta \equiv m_d - m_u \rightarrow 0$.  If we perform this calculation at multiple mass splittings $\delta$, we can extrapolate to the $\delta \rightarrow 0$ limit, at which point we can perform the necessary isospin rotations to relate the result to the difference between proton and neutron form factors.

\subsection{Form Factors}
In the isospin-broken case, a third form factor $F_3$ (which vanishes in the $\delta \rightarrow 0$ limit) contributes to the decomposition \cite{phiala}:
\begin{equation} \langle p_f | \mathcal O_\mu | p_i \rangle = \bar u(p_f) \left[ \gamma_\mu F_1(Q^2) + \frac{\sigma_{\mu\nu}q^\nu}{m_p+m_n} F_2(Q^2) + \frac{iq_\mu}{m_p+m_n} F_3(Q^2) \right] u(p_i) .
\label{form-factors}
\end{equation}
Determining the charge radius requires only $F_1$, so $F_2$ and $F_3$ are nuisance parameters.  

\subsection{Fitting the Data}
At a given value of $\delta$ and $Q^2$, we calculated the two-point functions $C_{2p}(t)$ and $C_{2n}(t)$ for the proton and neutron as well as the three-point function $C_3^\mu (t, \tau, \Gamma)$, where $t$ is the sink time, $\tau$ is the insertion time of the local vector current operator, $\mu$ is the index of the current inserted, and $\Gamma$ is the polarization matrix, which projects the initial and final baryons into given spin states (e.g. $\Gamma=\frac{1}{2}(1+\gamma_4)$ for unpolarized nucleons, $\frac{i}{2}(1+\gamma_4)\gamma_5\gamma_3$ for $z$-polarization).  We then formed the ratio
\begin{equation}
R_\Gamma^\mu (t,\tau) = \frac{C_3^\mu(t, \tau, \Gamma)}{C_{2f}(t)}\left[ 
\frac{C_{2f}(\tau)}{C_{2i}(\tau)}
\frac{C_{2f}(t)}{C_{2i}(t)}
\frac{C_{2i}(t-\tau)}{C_{2f}(t-\tau)}
\right]^{1/2} ,
\end{equation}
with $C_{2i,f}$ the initial- and final-state unpolarized 2-point functions, evaluated at momenta $\mathbf p_{i,f}$, respectively.  This ratio is designed to plateau to the matrix element in Eq.~(\ref{form-factors}), contracted with the appropriate $\Gamma$, for $(t-\tau), \tau \gg 1$ since factors exponentially decaying in $\tau$ and $t-\tau$ cancel.  By varying $\Gamma$ and $\mu$, we can extract the value of $F_1$ at different values of $t$ and $\tau$ for a given $\mathbf p_i, \mathbf p_f$.

When constructing the 3-point functions, we used a sequential source through the operator and varied the operator insertion time $\tau$ from 4 to 8 in lattice units.  
The magnitude of the momentum inserted at the operator was typically either 0 or 1 lattice units, and the magnitude of the sink momentum varied from 0 to 4 lattice units.  Momentum smearing \cite{momentum} was used to improve the signal of high-momentum states.  

We then performed a correlated multi-exponential fit to $F_1 (t,\tau)$ of the form $\alpha + \beta e^{-\Delta_1 (t-\tau)} + \gamma e^{-\Delta_2 \tau}$ and extract $\alpha$.  Note that, since the initial and final states have different masses, it is necessary to use different values $\Delta_{1,2}$ for the gap between the ground state and the first excited state in our fit.  

We computed $F_1$ at various nucleon mass splittings $\delta$ and momentum insertions $Q^2$, and we ultimately want $\left. \frac{\partial F_1}{\partial Q^2}\right|_{\delta=Q^2=0}$.  Thus, we fit our measurements of $F_1$ to the third-order polynomial 
\begin{equation}
F_1 (\delta, Q^2) = A (1 + C\delta^2 + D Q^2 + E \delta Q^2 + F \delta^3) ,
\label{f1-fit}
\end{equation} where $A$ is an overall renormalization constant (that differs from 1 since our current is not conserved), and we extract the parameter $D$ as our measurement of the slope.  We only used values of $Q^2$ with $|Q^2| \lesssim \delta^2$ so that this form of the fit is justified.  Note that the term linear in $\delta$ is omitted, as this vanishes by the Ademollo-Gatto theorem \cite{ademollo-gatto}.

\subsection{Lattice Parameters}
Since this was a preliminary study, we only used a single ensemble of gauge configurations. Repeating this work at lighter sea quark masses, larger volumes, and finer lattice spacing would be necessary to make predictions at the physical point.  We used a set of 750 $24^3 \times 64$ configurations generated with $N_f = 2+1$ flavors of clover fermions and a L\"uscher-Weisz gauge action with a lattice spacing $a \approx 0.12$ fm 
and a sea quark mass corresponding to $m_\pi \approx 450$ MeV.  We took the valence $u$ quark degenerate with the sea quarks and varied the mass of the $d$ quark (always keeping it at least as heavy as the $u$ quark) to obtain nucleon mass differences ranging from 0 to 0.44 GeV.  We computed correlators from 32 source locations per configuration at the smallest mass splitting; at larger mass splittings, where the computations were less noisy, we used fewer configurations.

\section{Results}
\begin{figure}
\centering
\begin{minipage}{0.47\textwidth}
\centering
\vspace{25pt}
\includegraphics[width=\textwidth]{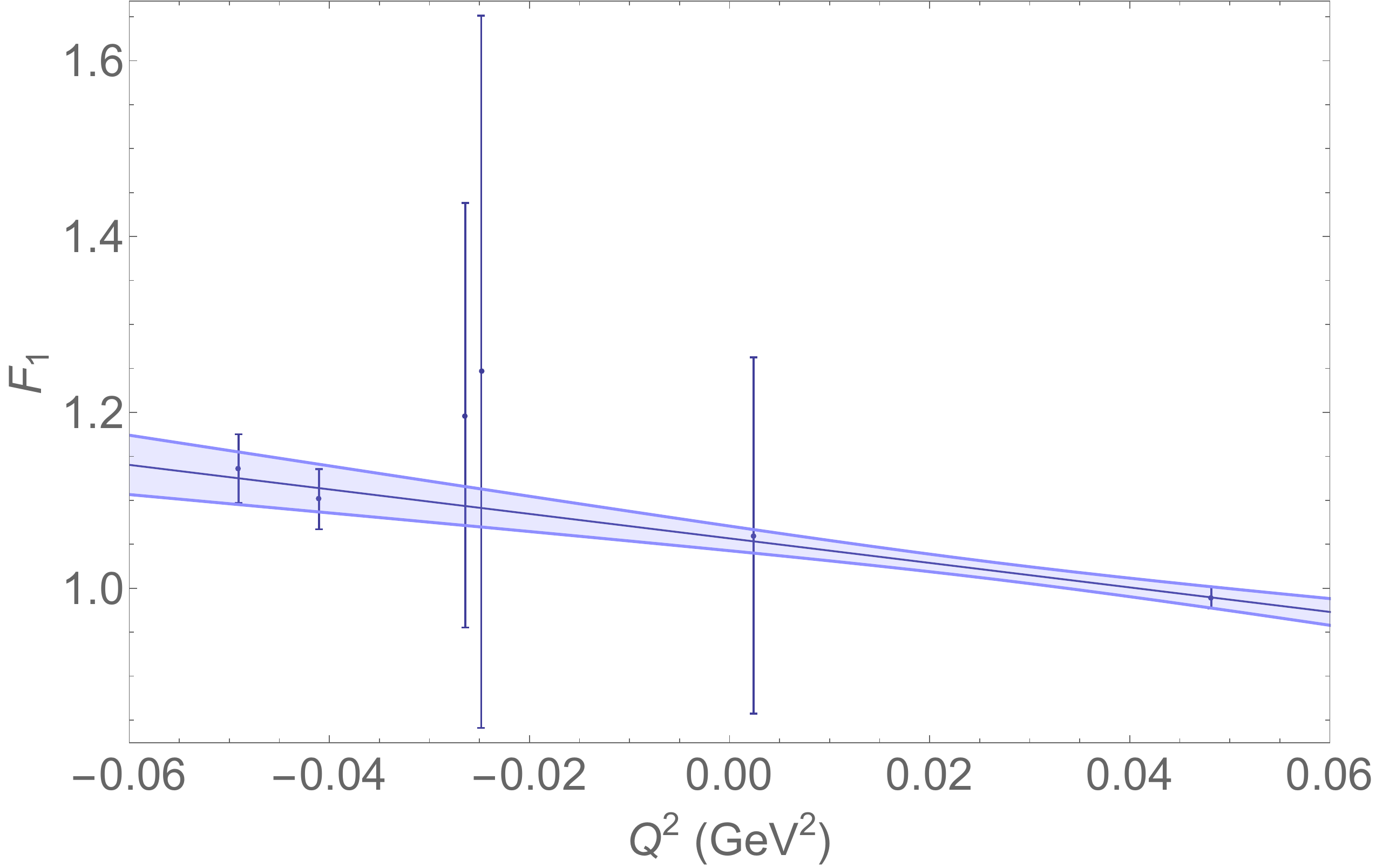}
\caption{The Dirac isovector form factor $F_1 (Q^2)$ at nucleon masses of $m_p = 1.52$ GeV, $m_n = 1.75$ GeV, the smallest mass splitting considered in this work.}
\label{1-splitting}
\end{minipage} \hfill
\begin{minipage}{0.47\textwidth}
\centering
\includegraphics[width=\textwidth]{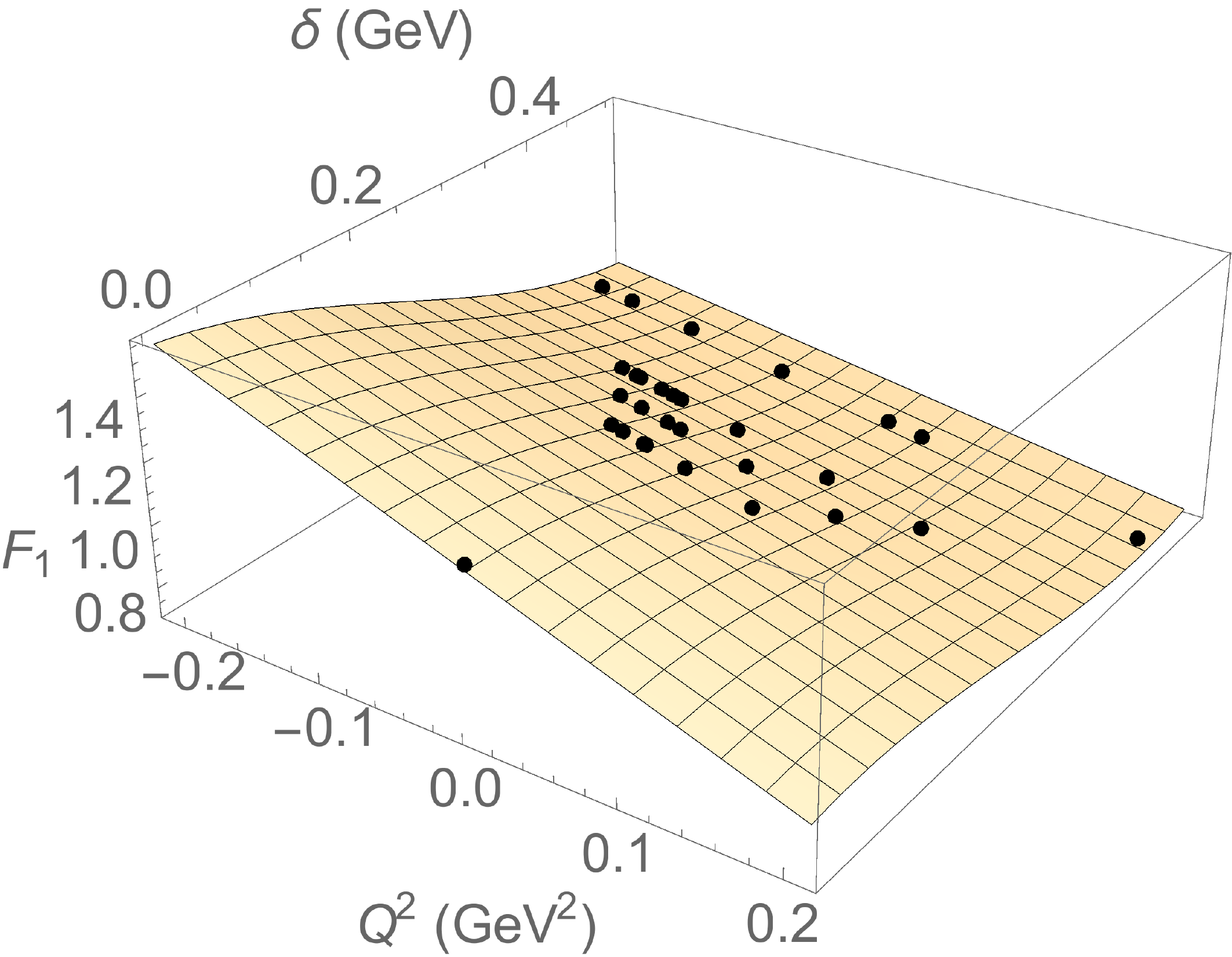}
\caption{The global fit surface $F_1 (Q^2, \delta)$ as well as the location of all data points in the $(Q^2, \delta)$ plane.  Note that we require $Q^2 \lesssim \delta^2$ so that the polynomial expansion (\ref{f1-fit}) will be valid.  
}
\label{all-data}
\end{minipage}
\end{figure}

At the smallest nonzero mass splitting ($m_p = 1.52$ GeV, $m_n = 1.75$ GeV), we varied the operator and sink momenta to determine $F_1(Q^2)$ (see Figure \ref{1-splitting}).  Repeating this process at various mass splittings allowed us to fit $F_1 (Q^2, \delta)$ (Figure \ref{all-data}) and extract the parameters in Eq.~(\ref{f1-fit}).
\vspace{-6pt}
\vspace{-2pt}

From the global fit shown in Figure \ref{all-data}, we extract $D = 1.37 \pm 0.32$ GeV$^{-2}$, giving a Dirac isovector charge radius squared of $0.320 \pm 0.074$ fm$^2$ at this pion mass of $m_\pi \approx 450$ MeV.  The uncertainty includes statistical fluctuations as well as systematic uncertainties in the fit ranges.  However, it neglects the uncertainty originating from truncating our fit function (\ref{f1-fit}) at third order.

The extracted charge radius is appreciably smaller than experimental measurements and lattice calculations at the physical point.  However, it appears consistent with other lattice calculations undertaken at unphysically heavy pion masses broadly comparable to the one considered here \cite{andrew-1}.

\section{Analysis}
As a preliminary study, our method allowed us to obtain an estimate of the isovector Dirac radius at quark mass corresponding to $m_\pi = 450$ MeV.  However, our method required a large amount of computational effort 
to obtain a fairly noisy signal at an unphysical pion mass.  We therefore discuss the prospects for improving our estimate and moving toward the physical point.

\subsection{Feasibility of the $\delta \rightarrow 0$ Extrapolation}
Most lattice measurements of the proton charge radius involve fitting the data with respect to $Q^2 \geq 0$ to some functional form and then computing the derivative of this function at $Q^2=0$.  Evaluating a function at the boundary of where it is fit to data introduces model dependence, especially since the closest data point to 0 is bounded away by one lattice unit of momentum.  (The $z$ expansion, discussed below, is model independent in principle but must be truncated.)  

By straddling $Q^2 = 0$ and including both timelike and spacelike momentum transfers, we can better constrain the behavior at $Q^2 = 0$.  However, this comes at the cost of introducing the mass splitting parameter $\delta$ that we then have to extrapolate to 0.  To reduce systematic errors associated with the $\delta \rightarrow 0$ extrapolation, we could take measurements at smaller $\delta$ than what was studied here, but this is hard for two reasons.  First, at a given $\delta$ with $|Q^2| < \delta^2$, the lever arm for determining the slope scales as $\delta^2$ and thus the statistical error scales as $\delta^{-2}$.  Compensating for this by increasing statistics means that the computational cost scales as $\delta^{-4}$.


Additionally, for $\delta \lesssim (2\pi/L)/\sqrt 2$, we also have the added complication that it is difficult to use nonzero momentum insertions (since one unit of momentum insertion would lead to $Q^2 > +\delta^2$ for reasonable values of $\mathbf p_f$).
Instead, we must perform the calculation using entirely timelike momentum transfers at these $\delta$.  If we fix $\mathbf q = 0$, then $Q^2 = -\delta^2$ if $\mathbf p_f = 0$ but $Q^2 \rightarrow 0$ as $|\mathbf p_f| \rightarrow \infty$.  By varying $\mathbf p_f$, we can in principle collect data over the entire range of timelike $Q^2$.  However, in practice, we are limited by the degradation of signal quality at large momenta.  Momentum smearing \cite{momentum} at both source and sink counteracts this partially, but using $|\mathbf p_f| \gtrsim m_N$ becomes impractical.  This restriction limits our calculations roughly to the range $Q^2 \in (-\delta^2, -\delta^2/2)$, one-quarter of the range available when 3-momentum insertion is valid.

In principle, we can take $\delta$ arbitrarily small to reduce systematic errors, na\"ively a positive aspect of our method.  However, practical considerations forced us to use values not much smaller than the minimum quantized momentum.
This seriously constrains the practicality of this method.


\subsection{Comparison to Other Methods}
In Ref.~\cite{andrew-1}, the Dirac isovector charge radius was computed at a variety of pion masses, the heaviest of which ($m_\pi \approx 350$ MeV) was broadly comparable to the one considered here.  Using the dipole fit ($F_1 = (1 + Q^2/\Lambda^2)^{-2}$), they obtain a similar uncertainty to ours but with far fewer propagator solves.  While they have uncontrolled systematics from the model-dependent dipole fit, it is difficult to compare the magnitude of this systematic to the $\delta \rightarrow 0$ extrapolations required here.

The $z$ expansion provides a method of extracting the slope at $Q^2 = 0$ from a wide range of spacelike $Q^2$ values.  Making use of a conformal map, the entire range of possible $Q^2$ of $(-4 m_\pi^2, \infty)$ is mapped to $z\in (-1, 1)$ so that a Taylor expansion around $Q^2 = 0$ can be performed.  In principle, the Taylor series is exact, but in practice, error is introduced when the series is truncated.  Varying the number of terms retained allows estimation of the approximation error.  In Ref.~\cite{andrew-2}, the $z$ expansion was used at the physical pion mass (where calculations are inherently noisier) to obtain an estimate of the isovector electric charge radius with an error bar comparable to ours.  Given that the $z$ expansion can be used to extract the form factor slope at a lighter pion mass with similar statistical uncertainty (using comparable statistics) and better-controlled systematics, the $z$ expansion appears to be preferable to our method.

\section{Conclusion}
At a pion mass of 450 MeV at a single lattice spacing, we computed the isovector Dirac radius squared to be $0.320 \pm 0.074$ fm$^2$ using a method that introduced a mass splitting between the neutron and proton.  While this method did allow us to perform a measurement of the Dirac radius, it proved to be noisy and computationally intensive, making a high-precision, physical-point measurement impractical.  This method is likely not competitive with other approaches for determining the proton charge radius.

\section{Acknowledgments}
W.~Detmold and A.~Grebe are supported by the U.S. Department of Energy under Early Career Research Award DE-SC0010495 and grant DE-SC0011090 and within the framework of the TMD Topical Collaboration of the U.S. Department of Energy, Office of Science, Office of Nuclear Physics, and by the SciDAC4 award DE-SC0018121. P.~E.~Shanahan is supported by the National Science Foundation under CAREER Award 1841699 and in part by Perimeter Institute for Theoretical Physics. Research at Perimeter Institute is supported by the Government of Canada through the Department of Innovation, Science and Economic Development and by the Province of Ontario through the Ministry of Research and Innovation.  We thank James Zanotti for helpful discussions.
We used the lattice QCD software Chroma and related packages \cite{chroma, quda, qdp-jit} for our calculations.


\begin{thebibliography}{99}
\bibitem{crema} A.~Antognini et al.~(CREMA collaboration), \emph{Proton Structure from the Measurement of 2S-2P Transition Frequencies of Muonic Hydrogen}, \emph{Science}, \textbf{339} (2013) 417--420. 
\bibitem{codata} P.~Mohr, D.~Newell, and B.~Taylor, \emph{CODATA Recommended Values of the Fundamental Physical Constants: 2014},  \emph{J. Phys. Chem. Ref. Data} \textbf{45} (2016) 043102 [atom-ph/1507.07956]. 
\bibitem{thomson} M.~Thomson, \emph{Modern Particle Physics}, Cambridge University Press, New York 2015.
\bibitem{peskin} M.~Peskin and D.~Schroeder, \emph{An Introduction to Quantum Field Theory}, Westview Press, Boca Raton 1995.
\bibitem{pdg} C.~Patrignani et al.~(Particle Data Group), \emph{\emph{N} Baryons ($S = 0$, $I = 1/2$)}, \emph{Chin. Phys. C}, \textbf{40} (2006) 100001.
\bibitem{andrew-1} J.~R.~Green et al., \emph{Nucleon Structure from Lattice QCD Using a Nearly Physical Pion Mass}, \emph{Phys. Lett. B} \textbf{734} (2014) 290--295 [hep-lat/1209.1687]. 
\bibitem{neutron} S.~Kopecky et al., \emph{Neutron Charge Radius Determined from the Energy Dependence of the Neutron Transmission of Liquid $^{208}$Pb and $^{209}$Bi}, \emph{Phys.~Rev.~C} \textbf{56} (1997) 2229--2237. 
\bibitem{andrew-2} N.~Hasan et al., \emph{Computing the Nucleon Charge and Axial Radii Directly at $Q^2 = 0$ in Lattice QCD}, \emph{Phys. Rev. D} \textbf{97} (2018) 034504 [hep-lat/1711.11385]. 
\bibitem{phiala} P.~Shanahan et al., \emph{SU(3) Breaking in Hyperon Transition Vector Form Factors}, \emph{Phys. Rev. D} \textbf{92} (2015) 074029 [hep-lat/1508.06923].  
\bibitem{momentum} G.~Bali et al., \emph{Novel Quark Smearing for Hadrons with High Momenta in Lattice QCD}, \emph{Phys. Rev. D} \textbf{93} (2016) 094515 [hep-lat/1602.05525]. 
\bibitem{ademollo-gatto} M.~Ademollo and R.~Gatto, \emph{Nonrenormalization Theorem for the Strangeness-Violating Vector Currents}, \emph{Phys. Rev. Lett.} \textbf{13} (1964) 264--266. 
\bibitem{chroma} R.~G.~Edwards (SciDAC and LHPC Collaboration), B.~Jo\'o (UKQCD Collaboration), \emph{The Chroma Software System for Lattice QCD}, in proceedings of \emph{Lattice 2004}, \emph{Nucl. Phys. B (Proc. Suppl.)}, \textbf{140} (2005) 832--834 [hep-lat/0409003]. 
\bibitem{quda} M.~A.~Clark et al., \emph{Solving Lattice QCD Systems of Equations Using Mixed Precision Solvers on GPUs}, \emph{Comput. Phys. Commun.}, \textbf{181} (2010) 1517--1528 [hep-lat/0911.3191]. 
\bibitem{qdp-jit} F.~Winter et al., \emph{A Framework for Lattice QCD Calculations on GPUs}, in proceedings of \emph{2014 IEEE 28th Intl. Parallel and Distributed Processing Symp.}, (2014) 1073--1082 [hep-lat/1408.5925]. 
\end{thebibliography}
\end{document}